\documentclass[superscriptaddress,nofootinbib,notitlepage,preprintnumbers]{revtex4}
\pdfoutput=1

\usepackage{graphicx}				
\usepackage{amssymb}
\usepackage{amsmath}								

\usepackage{color}

\usepackage[utf8]{inputenc}
\usepackage{longtable}
\usepackage{placeins}

\newcommand{\ras}[1]{{\color{blue}#1}}

\DeclareMathAlphabet \mathbfcal{OMS}{cmsy}{b}{n}

\begin{document}

\preprint{DESY 16-125}

\title{Bottom-Up Discrete Symmetries for Cabibbo Mixing}

\author{Ivo de Medeiros Varzielas}
\email{ivo.de@soton.ac.uk}
\affiliation{{\small School of Physics and Astronomy, University of Southampton,}\\
Southampton, SO17 1BJ, U.K.}
\author{Rasmus W. Rasmussen}
\email{rasmus.westphal.rasmussen@desy.de}
\affiliation{{\small Deutsches Elektronen-Synchrotron (DESY), Platanenallee 6, 15738 Zeuthen, Germany}}
\author{Jim Talbert}
\email{jim.talbert@physics.ox.ac.uk, james.talbert@desy.de}
\affiliation{{\small Rudolf Peierls Centre for Theoretical Physics,
University of Oxford,
1 Keble Road,}\\
Oxford,  OX1 3NP, U.K.}

\begin{abstract}We perform a bottom-up search for discrete non-Abelian symmetries capable of quantizing the Cabibbo angle that parameterizes CKM mixing.  Given a particular Abelian symmetry structure in the up and down sectors, we construct representations of the associated residual generators which explicitly depend on the degrees of freedom present in our effective mixing matrix.  We then discretize those degrees of freedom and utilize the \emph{Groups, Algorithms, Programming} (GAP) package to close the associated finite groups.  This short study is performed in the context of recent results indicating that, without resorting to special model-dependent corrections, no small-order finite group can simultaneously predict all four parameters of the three-generation CKM matrix and that only groups of $\mathcal{O}(10^{2})$ can predict the analogous parameters of the leptonic PMNS matrix, regardless of whether neutrinos are Dirac or Majorana particles.  Therefore a natural model of flavour might instead incorporate small(er) finite groups whose predictions for fermionic mixing are corrected via other mechanisms. 
\end{abstract}

\maketitle

\pagebreak
\section{Introduction}
Non-Abelian (NA) discrete flavour symmetries are powerful tools in the effort to explain the observed structure of fermionic masses and mixings.  In particular, they allow for precise predictions of mixing matrices and, when coupled with other auxiliary symmetries, can also help organize mass patterns.  Flavour models employing discrete symmetries are generically classified as `direct,' `semi-direct,' and `indirect' (see \cite{King:2013eh} for a review).  In the context of direct or semi-direct models, one might assume that, at very high energies normally at or above the GUT scale, a parent flavour symmetry $\mathcal{G_{F}}$ breaks to subgroups 
in the quark $\mathcal{G_{Q}}$ and lepton $\mathcal{G_{L}}$ sectors, which then subsequently break to subgroups in the charged lepton $\mathcal{G_{\text{e}}}$, neutrino $\mathcal{G_{\nu}}$, up $\mathcal{G_{\text{u}}}$ and down $\mathcal{G_{\text{d}}}$ sectors:
\begin{equation}
\label{eq:GF}
\mathcal{G_{F}}  \rightarrow \begin{cases}
				\mathcal{G_{L}}   \rightarrow \begin{cases} 
										\mathcal{G_{\nu}}
										\\
										\mathcal{G_{\text{e}}}
										\end{cases} \\
				\mathcal{G_{Q}} \rightarrow \begin{cases}
										\mathcal{G_{\text{u}}}
										\\
										\mathcal{G_{\text{d}}}
										\end{cases}
				\end{cases}
\end{equation}
This schematic simplifies if $\mathcal{G_{F}} = \mathcal{G_{L}} = \mathcal{G_{Q}}$, in which case the first arrow disappears and one only considers a single reduction to the final residual symmetries. If  $\mathcal{G_{L}}$ and  $\mathcal{G_{Q}}$ have separate origins,  $\mathcal{G_{F}}$ can be constructed from the direct product of the groups that give rise to $\mathcal{G_{L}}$ and  $\mathcal{G_{Q}}$.  Regardless of the breaking patterns, the parent symmetries must be NA in order for generations to be arranged in irreducible multiplets and\footnote{By considering all possible charge assignments in models with an Abelian $\mathcal{G_{F}} \sim Z_{n1} \times Z_{n2} \times ... $ in a Froggatt-Nielsen \cite{Froggatt:1978nt} scenario, as was done in \cite{Plentinger:2008up}, one can also achieve realistic mass and mixing relations.  However, these predictions are in terms of an unquantized texture parameter $\epsilon$, unlike the NA models we discuss here which fully predict the mixing matrix.}, similarly, the final pattern of residual symmetries present in the Standard Model (SM) Lagrangian must be Abelian and of order $N \geq $ number of generations (this requirement is due to the generations having distinct masses and non-trivial mixing).

Recently, the bulk of theoretical studies have focused on the leptonic sector, perhaps due to the flux of new experimental data indicating that the reactor angle $\theta_{13}$ is nonzero  \cite{An:2012eh} (see \cite{Forero:2014bxa, Gonzalez-Garcia:2015qrr, Capozzi:2016rtj} for global fits to neutrino mixing observables) and hence simple models based on, e.g., the flavour symmetry $A_{4}$
\cite{Ma:2001dn,Babu:2002dz,Altarelli:2005yp,Altarelli:2005yx,deMedeirosVarzielas:2005qg,Parattu:2010cy}
must be abandoned or substantially modified
\cite{Varzielas:2012ai,Ishimori:2012fg,Ahn:2013mva,Memenga:2013vc,Bhattacharya:2013mpa, Ferreira:2013oga,Felipe:2013vwa,Hernandez:2013dta,King:2013hj,Morisi:2013qna,Morisi:2013eca,Felipe:2013ie, Campos:2014lla,Hernandez:2015tna}.  Unfortunately, all model-independent scans of the lepton sector indicate that only large groups of $\mathcal{O}(10^{2})$ can quantize $\theta_{13}$ within 3$\sigma$, and even larger groups are needed to quantize the full PMNS matrix to a similar accuracy \cite{deAdelhartToorop:2011re, Lam:2012ga, Holthausen:2012wt, King:2013vna, Lavoura:2014kwa, Joshipura:2014pqa, Joshipura:2014qaa, King:2016pgv}. This statement is true for both Majorana and Dirac-type neutrinos, and regardless of whether the discrete flavour symmetry $\mathcal{G_{F}}$ is a subgroup of $SU(3)$ or $U(3)$, but only applies to direct models that completely predict the mixing angles.
This result was confirmed additionally  from general group theoretical arguments \cite{Fonseca:2014koa} and also from a bottom-up approach \cite{Talbert:2014bda} which we will discuss in detail below.

Furthermore,  studies addressing the quark sector are generally performed in light of the leptons. That is, people have searched for flavour symmetries in the quark sector \cite{Holthausen:2013vba, Ge:2014mpa, Ishimori:2014nxa, Yao:2015dwa} that have irreducible triplet representations or that can originate from the same groups that work for leptons (e.g. subgroups of $\Delta(6N^{2})$).
Inevitably, as one might predict given the extremely hierarchical structure of the CKM matrix, no finite group has been found that can predict all angles and phases of the CKM to any accuracy.  Small groups such as $D_{14}$  and other variants of the Dihedral family can predict the Cabibbo angle \cite{Blum:2007jz, Hagedorn:2012pg}, but still not within $3 \sigma$ \footnote{While smaller groups can produce viable leading order CKM matrices e.g. $S_3$ in \cite{Das:2015sca, Hernandez:2015dga}, the Cabibbo angle is not predicted in such models.}. Within this context,  it is prudent to consider the possibility that, if a NA discrete flavour symmetry does exist in nature, it is described by a small group whose predictions for fermionic mixing are modified, perhaps via Renormalization Group running \cite{JuarezWysozka:2002kx, Ross:2007az, Varzielas:2008jm} or additional symmetry breaking effects as have been studied for leptonic mixing \cite{Sierra:2013ypa, Hall:2013yha, Sierra:2014hea}. We adopt this philosophy in the present note, and focus on finite groups that can predict the Cabibbo angle at leading order.

We study Cabibbo mixing in the quark sector by utilizing the approach introduced in \cite{Talbert:2014bda}, which effectively inverts the arrows in Eq. (\ref{eq:GF}).  This method of `[re]constructing' finite flavour groups begins by identifying residual Abelian symmetries present in the Standard Model Yukawa sector and then building explicit representations of the generators of said symmetries.  By construction, those matrices depend on the same degrees of freedom present in the mixing matrices. One can then utilize the GAP system for computational finite algebra\footnote{http://www.gap-system.org/.  We use GAP4.7.} to close the groups generated by the representations. This approach essentially realizes an automation of the studies performed in \cite{Hernandez:2012sk, Araki:2013rkf, Lam:2009hn, Lam:2007ev}, and was previously applied to a special case of $\mu - \tau$ perturbed leptonic mixing. It is particularly useful as a model-building tool. The authors of \cite{Everett:2015oka} reiterated the bottom-up perspective in a non-automated fashion and also conceptually extended it to treat general CP symmetries.
 
This paper begins with a generic discussion of the residual discrete symmetries that are present in the quark mass sector of the SM in section \ref{sec:symmetries}.  Section \ref{sec:bottom-up} reviews the bottom-up method of \cite{Talbert:2014bda}, and elaborates the specific details of this implementation.  Results are presented in section \ref{sec:results} before we discuss the trends and limitations of our search method in section \ref{sec:Limits}.  We give closing remarks in section \ref{sec:conclusions}.  

\section{The symmetries of the quark Yukawa sector \label{sec:symmetries}}
The philosophy of building NA flavour symmetries via the identification of residual Abelian symmetries present in the SM Lagrangian has been approached both analytically and numerically over the last couple of years (cf. references above).  We adopt this philosophy below to identify the relevant residual symmetries $\mathcal{G_{\text{u}}}$ and $\mathcal{G_{\text{d}}}$ of the quark mass sector, closely following the discussion and notation of \cite{Araki:2013rkf}.

The SM Lagrangian for quark masses is given by:
\begin{equation}
\label{eq:SML}
\\
-\mathcal{L} = \bar{U}_{R} \hat{M}_{U} U_{L} + \bar{D}_{R}M_{D}D_{L} + h.c.
\\
\end{equation}
where $U_{L,R} \equiv (u, c, t)^{T}_{L,R}$, $D_{L,R} \equiv (d, s, b)^{T}_{L,R}$ and $\hat{M}_{U} \equiv diag\lbrace m_{u}, m_{c}, m_{t}\rbrace$.  Hence we are in the basis where the up quarks are diagonal. It is clear from Eq. (\ref{eq:SML}) that the Lagrangian is invariant under the action of a U(1) symmetry for each active generation and, noting that $U_L$ and $D_L$ belong to the same $SU(2)_L$ doublet, the natural residual symmetry of both up and down quark mass terms is $U(1)^{3}$.  We are currently only interested in discrete flavour symmetries, so we focus on discrete cyclic subgroups and their direct products:
\begin{equation}
\label{eq:GQ}
\mathcal{G_{Q}}  \rightarrow  \begin{cases}
				\mathcal{G_{\textit{u}}} \sim Z_{n}^{u},\,  Z_{n1}^{u} \times Z_{n2}^{u}
				\\
				\mathcal{G_{\textit{d}}} \sim Z_{m}^{d}, \, Z_{m1}^{d} \times Z_{m2}^{d}
				\end{cases}
\end{equation}
We assign $\mathcal{G_{\textit{u}/\textit{d}}}$ to a single cyclic $Z_{n/m}$ (with ($n,m$) the order of the associated generator) in analogy to the usual choice made for the charged leptons, or to a direct product group in analogy to the maximal $Z_{2} \times Z_{2}$ symmetry that exists for Majorana neutrino mass matrices (see e.g. \cite{King:2013eh}).  However, in this case our cyclic generators are of course not bound to be of order two in either the up or down sectors.  Denoting the generator(s) of $Z_{u}$ as $T_{l}$ and the generator(s) of $Z_{d}$ as $S_{Di}$, the actions of the above residual symmetries on the left-handed fields that are relevant for mixing are represented by  \footnote{The transformation properties on the right-handed fields are not important for this puprose, as there is no physical right-handed mixing in the SM.}:
\begin{align}
U_{L} &\rightarrow T_{l} U_{L}\\
D_{L} &\rightarrow S_{Di}D_{L}
\end{align}
where for three generations
\begin{equation}
\label{eq:T}
T_{l} = diag\left(e^{i \Phi_{1}}, e^{i \Phi_{2}}, e^{i \Phi_{3}}\right)_{l}
\text{  where  } 
\Phi_{j} = 2\pi \frac{\phi_{j}}{n} \text{   }
\end{equation}
Both $\phi_{j}$ and $n$ are integers, with $n$ representing the order of the generator.  In the down sector, $S_{Di}$ are given as the rotated generators that depend on the explicit degrees of freedom present in the unitary mixing matrix:
\begin{equation}
\label{eq:SD}
S_{Di}(\lbrace \Theta_{k}, \alpha_{j}\rbrace) = U_{CKM}(\Theta_{k}) \,S(\alpha_{j})_{i}\,U^{\dagger}_{CKM}(\Theta_{k})
\end{equation}
where $S_{i}$ are diagonal matrices analogous to Eq. (\ref{eq:T}) with phases $\alpha_{j}$ and $\lbrace\Theta_{k}\rbrace$ are whatever mixing angles and CP violating phases are present in $U_{CKM}$.

If we wish to assume that $\mathcal{G}_{u}, \mathcal{G}_{d} \subset SU(3)$ (or SU(2) for the limiting case of LO Cabibbo mixing) we can of course impose charge constraints on $T$ and $S_{D}$ such that:
\begin{equation}
\sum_{j} \Phi_{j},\alpha_{j} \equiv 0 \textit{     }mod \textit{   }2 \pi
\end{equation}
However, in this study we make no such constraint and thus look at the relevant U(3)/U(2) groups to be less restrictive.  
\section{A bottom-up technique for closing groups \label{sec:bottom-up}}
In this section we first briefly review our procedure for finding phenomenologically viable NA discrete symmetries (though we refer the reader to \cite{Talbert:2014bda} for a more detailed discussion) before outlining some of the specific details of its application in the quark sector.
\subsection{[Re]constructing finite flavour groups}
Having assigned residual Abelian discrete symmetries to the up and down sectors, one is in a position to search for NA symmetries in a bottom-up manner by examining the possible groups closed by the combination of their associated `residual' generators.  To this end, the basic maneuvers executed by our scripts can be summarized as follows:
\begin{enumerate}
\item {\bf{Discretization:}} By construction our generators depend on the same degrees of freedom $\lbrace \Theta_{k} \rbrace$ present in the mixing matrices under consideration, which are of course continuous.  In order to find finite flavour groups one must impose a discretization on $\lbrace \Theta_{k} \rbrace$.  Two examples to this end are given by:
\begin{subequations}\label{subeq:par}
\begin{align}
\label{subeq:par1}
\tan(\Theta_{k}) &= \sqrt{\frac{c}{1-c}}\\
\label{subeq:par2}
\Theta_{k} &= c\pi
\end{align}
\end{subequations}
where $c \equiv \frac{a}{b}$ and $(a,b) \in Integers$.  In the first scheme there is only the single parameter $c$ ($c \in \left[0,1\right)$) because we have restricted ourselves to the unit circle. The second scheme rationally discretizes the angle $\Theta$ itself, where we insist that $\Theta$ lie between 0 and $2\pi$ ($a \leq 2b$) to avoid any degeneracy\footnote{As it turns out, systematic studies of finite subgroups of $SU(3)$ \cite{Yao:2015dwa} show that Eq. (\ref{subeq:par2}) is a rather comprehensive scheme for discretizing the possible mixing angles for both quarks and Dirac neutrinos. Therefore will we only consider Eq. (\ref{subeq:par2}) in this study. Note however at least one relevant counter-example is the canonical tri-bimaximal mixing form, which would require a discretization along the lines of Eq. (\ref{subeq:par1}), with $c = \frac{1}{3}$ \cite{Talbert:2014bda}.}. Our `scans' are then implemented over the variables $(a,b)$ --- as the range of the examined $(a,b)$ parameter space grows, so does the number of generators we construct and by proxy the number of potential finite groups we can close.  
\item  {\bf{Experimental Constraints:}}  We can look to experimental constraints to limit/tune the (a,b) search space in order to find groups that quantize phenomenologically relevant mixing parameters.  Experimental data is often presented with respect to the PDG parameterization of $U_{CKM}$, and hence the most general constraint we can make on any matrix element is given by:
\begin{equation}
\label{eq:equate2}
\parallel U^{PDG}_{min} \parallel ^{2} \text{  }  \leqq \text{  }  \parallel U_{ij}\left(c_{k}\right) \parallel^{2} \text{  }  \leqq \text{  }  \parallel U^{PDG}_{max} \parallel ^{2}
\end{equation}
In words, we can insist that no matrix element be any greater/smaller than the largest/smallest (experimentally determined) elements of $U_{PDG}$. We have inequalities in Eq. (\ref{eq:equate2}) as opposed to equalities because the finite representations of the residual symmetries know nothing of the ordering of rows and columns (this constraint is relatively generic because while the discrete group can predict entries in the mixing matrix, it can not predict their placement). Thus while we can constrain \textit{matrix elements} within $x\mathcal{\sigma}$ (where $x$ is an arbitrary integer), we may not immediately predict the mixing angles to within $x\mathcal{\sigma}$, and require an additional cut at the end of the search. One may of course impose Eq. \eqref{eq:equate2} on multiple elements of a given class. A common practice in the leptonic sector is to assign the smallest predicted entry of the mixing matrix to the (13) entry, which is the smallest according to observation. However, in cases where one attempts to predict a specific $2 \times 2$ submatrix, as we intend to do for quarks, we can choose to place the smaller entries in the off-diagonal elements.
\item {\bf{GAP Implementation:}}  We wish to build explicit representations of $T_{l}$ and $S_{Di}$ in GAP, and hence we must translate our parameterizations of $\lbrace \Theta_{k} \rbrace$ given by Eq. (\ref{subeq:par}) into GAP objects.  For the second discretization scheme, Eq. (\ref{subeq:par2}), this amounts to creating lists of the following:
\begin{subequations}
\label{subeq:GAP}
\begin{align}
\label{subeq:GAP2a}
\cos \left(c\right) &= \frac{E\left(2b\right)^{a} + E\left(2b\right)^{-a}}{2}\\
\label{subeq:GAP2b}
\sin \left(c\right) &= \frac{E\left(2b\right)^{a} - E\left(2b\right)^{-a}}{2 E\left(4\right)}
\end{align}
\end{subequations}
where $E$ returns the primitive N-th root of unity, $E\left(N\right)\equiv e^{\frac{2 \pi i}{N}}$. Additional parameterizations would obviously lead to more variants of Eq. (\ref{subeq:GAP})\footnote{For example, were we to also consider the first scheme Eq. (\ref{subeq:par1}), the corresponding GAP objects would look like:
\begin{align*}
\cos \left(\Theta\left(c\right)\right) &= ER\left(1-\frac{a}{b}\right)
\\
\sin \left(\Theta\left(c\right)\right) &= ER\left(\frac{a}{b}\right)
\end{align*}
where $ER$ is a square root operation for a rational number $N$, $\sqrt{N}$.}.
\item {\bf{Generator Formation:}}  Form the explicit representations of viable $S_{D i}$ Eq.~(\ref{eq:SD}) and $T_{l}$ Eq.~(\ref{eq:T}) via Eq.~\eqref{subeq:GAP}.
\item {\bf{Close the Groups:}}  Now that we have the relevant GAP representations of $S_{D i}$ and $T_{l}$ in a specified interval of $\left(a,b,\phi_{j},\alpha_{j}, n, m\right)$ and a user-determined experimental $\sigma$-range, we wish to close the groups  $\mathcal{G_{F/Q/L}}$ generated by them.  GAP is capable of constructing groups directly from matrix representations of generators using the $GroupWithGenerators$ command. As we will discuss below, for instance, the idea of this paper is to form all groups closed by
\begin{subequations}\
\label{subeq:groups}
\begin{align}
\label{subeq:full4}
\mathcal{G_{Q}} &= \lbrace S_{D1}, S_{D2}, T_{1}, T_{2}  \rbrace
\\
\label{subeq:full1}
\mathcal{G_{Q}} &= \lbrace S_{D1}, S_{D2}, T \rbrace
\\
\label{subeq:full2}
\mathcal{G_{Q}} &= \lbrace S_{D}, T_{1}, T_{2} \rbrace
\\
\label{subeq:half}
\mathcal{G_{Q}} &= \lbrace S_{D}, T \rbrace 
\end{align}
\end{subequations}
Eq. \eqref{subeq:full1} treats the case where $\mathcal{G}_{d} \sim Z_{m1}^{d} \times Z_{m2}^{d}$ and $\mathcal{G}_{u} \sim Z_{n}^{u}$ whereas Eq. \eqref{subeq:full2} treats the case where $\mathcal{G}_{d} \sim Z_{m}^{d}$ and $\mathcal{G}_{u} \sim Z_{n1}^{u} \times Z_{n2}^{u}$, and so on.
\item {\bf{Analyze:}} Not all groups closed will be finite, of small-order, NA, etc.  GAP contains a host of internal commands that, given a group structure, can be used to filter results based on user-defined preferences.  For our purposes we are only concerned with small(ish), finite, NA groups.  We impose cuts to that end (details below), and then identify the remaining flavour symmetry candidates with the $GroupID$ and $StructureDescription$ commands\footnote{Note that $StructureDescription$ is \emph{not} an isomorphism invariant command --- two non-isomorphic groups can return the same group structure string while isomorphic groups in different representations can return different strings.  It is designed to primarily study small groups of $\mathcal{O}(\mathcal{G}) \lesssim 100$, which is our principal goal in this study.  On the other hand, the $GroupID$ is unique.}.   Our scripts carefully keep track of the parameters $\lbrace a,b,\alpha_{j}...\rbrace$  associated to the final group structure we identify, so that we might have explicit information on the representations of the residual generators, which is of course relevant at the model-building stage.
\end{enumerate}
\subsection{Specific details of this study}
The CKM mixing matrix is given in the Wolfenstein parameterization \cite{Agashe:2014kda} by:
\begin{equation}
\label{eq:Wolf}
U_{CKM} =
\left(
\begin{array}{ccc}
 1- \lambda^{2}/2 & \lambda & A \lambda^{3} (\rho - i \eta)\\
 -\lambda & 1- \lambda^{2}/2  & A \lambda^{2}\\
 A \lambda^{3} (1 - \rho - i \eta) & -A \lambda^{2} & 1\\
\end{array}
\right) + \mathcal{O}(\lambda^4)
\end{equation}
Since $\lambda = .22537^{+.00061}_{-.00061}$, $A = .814^{+.023}_{-.024}$, $\bar{\rho} = .117^{+.021}_{-.021}$ and $\bar{\eta} = .353^{+.013}_{-.013}$ \cite{Agashe:2014kda} (where $\bar{\rho} = \rho (1 - \lambda^2/2+...)$ and $\bar{\eta} = \eta (1 - \lambda^2/2+...)$) \cite{Buras:1994ec}, we find
\begin{equation}
\label{eq:num}
|U_{CKM}| \simeq
\renewcommand{\arraystretch}{1.4}
\left(
\begin{array}{ccc}
  \left(.97441 \atop .97413 \right)  &  \left(.22597 \atop .22475 \right)  & \left(.00370 \atop .00340 \right)  \\
  \left(.22583 \atop .22461 \right) &  \left(.97358 \atop .97328 \right)  &  \left(.0426 \atop .0402 \right) \\
  \left(.00919 \atop .00854 \right) &  \left(.0416 \atop .0393 \right)  &  \left(.99919 \atop .99909 \right) \\
\end{array}
\right)
\end{equation}
The hierarchical nature of the quark mixing matrix is now obvious; exterior off-diagonal elements are suppressed by one to two orders of magnitude and the upper 2 $\times$ 2 sub-matrix very nearly approximates an $SO(2)$ rotation about the Cabibbo angle: 
\begin{equation}
\label{eq:LOCKM}
U^{LO}_{CKM} \simeq
\left(
\begin{array}{cc}
\cos\theta_{C} & \sin\theta_{C}\\
-\sin\theta_{C} & \cos\theta_{C}\\
\end{array}
\right)
\end{equation}
Such a matrix does not exhibit CP violation.  Considering the numerical values of Eq. (\ref{eq:num}) and the fact that no discrete group has been found that quantizes them, it makes sense to study only Eq. (\ref{eq:LOCKM}) with the bottom-up technique described above.  Given the symmetry assignments of Eq. (\ref{eq:GQ}) and Eq. (\ref{eq:T}), we find explicit forms for the effective 2-generation $S_{Di}$.
\begin{equation}
\label{eq:SD1}
S_{Di}=
\renewcommand{\arraystretch}{1.5}
\left(
\begin{array}{cc}
e^{i \alpha_{1i}}\cos^{2}\theta_{C} + e^{i\alpha_{2i}} \sin^{2}\theta_{C} & \left(e^{i\alpha_{2i}}-e^{i\alpha_{1i}}\right)\cos\theta_{C}\sin\theta_{C} \\
\left(e^{i\alpha_{2i}}-e^{i\alpha_{1i}}\right)\cos\theta_{C}\sin\theta_{C} & e^{i \alpha_{2i}}\cos^{2}\theta_{C} + e^{i\alpha_{1i}} \sin^{2}\theta_{C}
\end{array}
\right)
\end{equation}
\\
In the event that $\mathcal{G}_{d} \sim Z_{m}^{d}$ and not a direct product, the index $i$ is meaningless.  We clearly then have 3-5 degrees of freedom that need to be discretized in the down sector via Eq. (\ref{subeq:par2}), $\lbrace \alpha_{1i}, \alpha_{2i}, \theta_{C} \rbrace$, and of course 2-4 degrees of freedom in the up sector, $\lbrace \Phi_{1l}, \Phi_{2l} \rbrace$.  Then, for each physical degree of freedom discretized using Eq. (\ref{subeq:par2}), there are two corresponding integers $c=\frac{a}{b}$ which must be scanned over in the bottom-up approach.   For all phases $\alpha$ and $\Phi$ we restrict $a \in \lbrace-1,0,1\rbrace$ and $b \in \lbrace2...\text{Max}(\mathcal{O}(T_{l},S_{i}))\rbrace$ where b, for diagonal matrices, also represents the order of the generator.  It must be at least two so that generations can be distinguished, and its maximum value is user-defined and specified below for various scans.  We vary both the discretization parameter ranges and allowed quantization range associated to the physical mixing angle $\theta_{C}$ in each scan.

\section{Results \label{sec:results}}
In this section we present our results and some discussion given the four assignments for the residual symmetries $\mathcal{G_{\textit{d}/\textit{u}}}$. In each subsection we reference tabled results of the groups found when searching within the parameter ranges discussed above and/or below.  The first column of each table gives the parameter $c$, which is a direct proxy for the Cabibbo angle.  The following 2-3 columns give the diagonal entries of the $2 \times 2$
matrix representations of $T_{l}$ and $S_{i}$, as discussed in Section \ref{sec:symmetries}.  There are three columns when either $\mathcal{G}_{u}$ or $\mathcal{G}_{d}$ is a direct product.  The fourth (fifth) column gives the unique ID of the given group closed as labeled in the GAP system, and the following column the associated group structure as given by the $StructureDescription$ command.  $D_{N}$ corresponds to the Dihedral group of order N, $Q_{N}$ to quarternions of order N, and $QD_{N}$ to Quasi-Dihedrals of order N.  We also remind the reader of the isomorphism structure of $\Sigma(2N^{2})$ groups (see Appendix \ref{App}),
\begin{equation}
\Sigma (2N^{2}) \equiv (Z_{N} \times Z_{N^{\prime}}) \rtimes Z_{2}
\end{equation}
and for simplicity we have also arbitrarily named the following groups:
\begin{equation*}
\Psi(N,M) \equiv (Z_{N} \times Z_{M}) \rtimes Z_{2}
\end{equation*}
Finally, the last column gives the value for $\sin \theta_{C}$ quantized by the group.  In all tables we only present results with non-trivial charge reassignments in the residual symmetry generators and non-trivial permutations of the parameter $c$.  That is, we do not show results where the same group quantizes the same mixing matrix, but with different diagonal matrix elements in $T_{l}$ or $S_{i}$, or results with explicitly different $c$ but equivalent $\sin(c \pi)$.

\subsection{\boldmath{$G_{d} \sim Z_{m}^{d}$, \, $G_{u} \sim Z_{n}^{u}$}}
\label{subsec:ud}

We begin by assigning a single cyclic symmetry to both the up and down sectors, which reflects the simplest possible discrete symmetry scenario, and scanning over the possible NA finite groups closed with the associated generator representations.  We present the scan results in Tables \ref{TS_wide} and \ref{TS_tight}, which are also discussed in more detail here than in following sections.

In Table \ref{TS_wide} we allow for a rather large window for the Cabibbo angle, $.2 \leq \sin \theta_{C} \leq .3$ to be sure we actually had predictions in our first simulation, and restrict the discretization parameters to $a,b \in \lbrace 0,1 ...50 \rbrace$, choices which when combined yield 52 values of the parameter $c$.  The order of the residual generators is restricted to $\mathcal{O}(T,S) \leq 4$, which (given the choices for the phase parameters described above) yields 19 unique diagonal generators to be distributed to both the up and down sectors.  This means there are $19\cdot52=988$ unique non-diagonal generators $S_{D}$ in the down-sector and $19^{2}\cdot52=18772$ different combinations of generators that could potentially close NA finite groups.  To quicken the scans, we first confirm that $\mathcal{O}(S_{D}\cdot T) < \infty$\footnote{We test all such combinations for other symmetry assignments where more generators are considered.}, as will be the case for any finite group generated by $S_{D}$ and $T$. Then, as our stated goal is to primarily search for small flavour groups, we restrict the order of the parent group to $\mathcal{O}(\mathcal{G}_{\mathcal{Q}}) \leq 75$.  Table  \ref{TS_wide} gives our results given these `bottom-up' inputs. In this and other Tables containing results one can clearly identify some cases that are subgroups of other groups listed. One sees that a host of group structures are obtained with $D_{14}$, $D_{28}$ and  $Z_{7} \rtimes Z_{4}$ providing the best prediction of $\sin \theta_{C} \simeq .2225$ ($c=1/14$).  The Dihedral groups $D_{n}$ and $D_{2n}$ predict the same Cabibbo angle because the order is linked to an integer multiple of the denominator of the input parameter $c$ for the groups.  We find that other semi-direct products, $\Psi$, and $Q$ groups all predict less interesting values for $\sin \theta_{C}$.

One may also observe that $D_{46}$ is generated for different Cabibbo angles ($\sin \theta_C \simeq 0.2698$ and $\sin \theta_C \simeq 0.2035$). Similarly $D_{62}$ is also generated for different Cabibbo angles ($\sin \theta_C \simeq 0.2013$ and $\sin \theta_C \simeq 0.2994$).  This can be understood in the following way: Each Dihedral group of order $2n$ is capable of predicting $n$ angles depending on which subgroups are left as residual symmetries. For the particular cases of $2n=46$ and $2n=62$ there would be $23$ and $31$ predictions available, and it just happened that two of those were within the $.2 \leq \sin \theta_{C} \leq .3$ window we allowed.

Indeed, with $Z_2$ residuals and 2-d irreps., we consistently reconstruct Dihedral groups of an order related to the denominator $b$ of the discretization parameter $c$, and see that the 2-d irrep. has a familiar geometrical interpretation. In detail, with $\theta_C = \frac{a}{b} \pi$, we reconstruct a $2\times2$ element $g_{rot}$ of the group with determinant $det(g_{rot})=1$ (usually this can be $T S_D$, if both $T$ and $S_D$ have determinant $-1$ as they do in the cases in Table \ref{TS_wide}, where they are both geometrical reflections). $g_{rot}$ can be seen, through the use of trigonometric identities, to be a geometrical rotation by an angle that is an integer multiple of $\frac{1}{b} \pi$ of order $n$: $g_{rot}^n$ is the identity. The $n$ elements that are rotations (with positive determinant) can be obtained by taking the powers of $g_{rot}$. The other $n$ elements of the Dihedral group are reflections (with negative determinant) and can be obtained by multiplying each of the distinct rotations by one of the reflections.

These results can be compared to Table \ref{TS_tight}, where we tighten the Cabibbo window to $.22414 \leq \sin \theta_{C} \leq .22658$ (in closer accordance to the PDG allowed experimental range) while simultaneously broadening the discretization range to $a,b \in \lbrace 0,1 ...100 \rbrace$.  We now only find 8 allowable values for $c$ ranging from $\frac{4}{55}$ to $\frac{90}{97}$.  We further restrict $\mathcal{O}(T,S) \leq 3$ and $\mathcal{O}(\mathcal{G}_{\mathcal{Q}}) \leq 1000$, as such values for $c$ will intuitively generate much larger groups than before. Indeed, we now find only larger Dihedral groups with the smallest ones being $D_{110}$ and $D_{138}$.  However, these groups obviously yield better predictions for $\sin \theta_{C}$ --- all groups except $D_{110}$ and $D_{220}$ showing up in Table \ref{TS_tight} predict angles that fall within the PDG allowed ranges in Eq. (\ref{eq:num}).  Were we to allow for an even finer gridding of $a$/$b$, we should expect to be able to find Dihedral groups predicting ever more precise mixing angles. 

$D_{14}$ and other Dihedral groups have been known in the literature for some time \cite{Blum:2007jz, Hagedorn:2012pg}.  Our approach reveals how generating them is nearly a trivial matter.  Consider the original mixing matrix Eq. (\ref{eq:LOCKM}), which represents an $SO(2)$ rotation in the Cabibbo plane. This can obviously be thought of as a circle, and quantizing $\theta_{C}$ to a rational multiple of $\pi$ corresponds to carving regular polygons out of said circle.  Dihedral groups encode the symmetries of polygons ($D_{8}$ is the symmetry of a square, e.g.), so it is no surprise that they show up throughout our scans.  It is also no surprise that a finer gridding in the discretization parameters generates larger groups;  the number of sides of the associated polygons increases. Given the order of the Dihedral groups $D_{110}$ and $D_{138}$ found in Table \ref{TS_tight}, we restrict $\mathcal{O}(G_{F}) \lesssim 75$, since a Dihedral group will always be able to trivially predict a mixing angle in a given range, if the order is high enough. Also note that Dihedral groups are not found in more universal, top-down scans like that in \cite{Yao:2015dwa} because most such studies insist that $\mathcal{G}_{\mathcal{Q}}$ contain 3-D irreducible representations.

Given the final results in Tables \ref{TS_wide} and \ref{TS_tight}, one can then directly reconstruct the explicit generator representations (in an appropriate basis) that work for realistic direct and semi-direct discrete models of flavour. As an example, consider line 13 of Table \ref{TS_wide}, where we immediately read off that that the numerical mixing matrix
\begin{equation}
\label{eq:mixexample}
U^{LO}_{CKM} \simeq
\left(
\begin{array}{cc}
.974928 & .222521\\
-.222521& .974928\\
\end{array}
\right)
\end{equation}
is predicted from the NA finite group $Z_{7} \rtimes Z_{4}$ ($SmallGroup(28,1)$) generated by the following explicit matrix representations in the up and down sectors: 
\begin{equation}
\label{eq:S74}
T^{(28,1)} = 
\renewcommand{\arraystretch}{1.5}
\left(
\begin{array}{cc}
-i & 0\\
0 & i\\
\end{array}
\right)
\,\,\,\,\,\,\,\,\,\,\,\,\,\,
S^{(28,1)}_{D}=
\renewcommand{\arraystretch}{1.5}
\left(
\begin{array}{cc}
-i \cos \frac{2\pi}{14} & i \sin \frac{2 \pi}{14}\\
i \sin \frac{2 \pi}{14} & i \cos \frac{2\pi}{14}\\
\end{array}
\right)
\end{equation} 
where we made use of some trigonometric identities. In direct and semi-direct flavour models, the vacuum expectation values of various `flavons' must be invariant under the operation of the group elements corresponding to these matrices (so that the broken family symmetry reproduces the data at the level of the Standard Model Lagrangian Eq. (\ref{eq:SML})).  Hence the bottom-up method can be particularly useful for model-builders.

As a final note, the familiar reader may question why Tables \ref{TS_wide} and \ref{TS_tight} do not contain a greater diversity of group structures\footnote{In both Tables \ref{TS_wide} and \ref{TS_tight} we have restricted the $\mathcal{O}(T,S)$ to the same maximum value. One may wonder whether more interesting structures can be found by allowing one subgroup to have a larger maximum order.  We have performed a scan along these lines where  $\mathcal{O}(T)  \leq 6$ but $\mathcal{O}(S)  \leq 4$.  We again put $a,b \in \lbrace 0,1 ...50 \rbrace$, $.2 \leq \sin \theta_{C} \leq .24$, and restrict $(\mathcal{G}_{\mathcal{Q}}) \leq 75$.  With these inputs we find no new group structures and no new predictions for the Cabibbo angle.}. For example, it is well-known that $A_{4}$, the alternating symmetry of the tetrahedron, has been used to predict unit (i.e. trivial) mixing in the quark sector \cite{Ma:2001dn,Babu:2002dz,Altarelli:2005yp,Altarelli:2005yx,deMedeirosVarzielas:2005qg}\footnote{This is expected from other groups, when both sectors are broken to the same subgroup.}, which may be a reasonable first-order approximation to $U_{CKM}$.  Yet it is clear that we will never obtain this prediction with our approach.  Unit mixing translates to a diagonal down-sector generator Eq. (\ref{eq:SD}), which when combined with the diagonal up-sector generator Eq. (\ref{eq:T}) will never close a NA finite group, regardless of the associated charges --- diagonal matrices commute.  As another example, consider $S_{3}$, the symmetry group of the triangle. We do not find it in Tables \ref{TS_wide} or \ref{TS_tight}, but it has a single two-dimensional irreducible representation, that can be generated by two matrices that fit into the forms of Eq. (\ref{eq:T}) and Eq. (\ref{eq:SD1}) (the two-dimensional form of Eq. (\ref{eq:SD})).  This absence is due to limits we put on the Cabibbo quantization window --- $S_{3}$ predicts a much larger value for $\sin \theta_{C}$ than .3 (.7071).  In Appendix \ref{App} we look at non-physical values of the Cabibbo angle and show that, indeed, many other group structures can be found using our method.  In Section \ref{sec:Limits} we briefly discuss the sensitivity of the method to user-defined parameter choices.
\begin{table}
\begin{tabular}{|c|c|c|c|c|c|}
\hline 
 c & $T_{diag}$ & $S_{i}$ & GAP-ID & Group Structure & $\sin \theta_{C}$ \\
\hline 
$\frac{1}{11}$ & [-1, 1] & [-1, 1] & [22, 1] & $D_{22}$ & $.2817$\\
$\frac{1}{11}$ & [1, -1] & [-1, 1] & [44, 3] & $D_{44}$ & $.2817$\\
$\frac{1}{11}$ & [-i, i] & [-i, i] & [44, 1] & $Z_{11} \rtimes Z_{4}$ & $.2817$\\
$\frac{1}{12}$ & [-1, 1] & [-1, 1] & [24, 6] & $D_{24}$ & $.2588$\\
$\frac{1}{12}$ & [-i, i] & [-1, 1] & [24, 8] & $\Psi(6,2)$ & $.2588$\\
$\frac{1}{12}$ & [-i, i] & [-i, i] & [24, 4] & $Z_{3} \rtimes Q_{8}$ & $.2588$\\
$\frac{1}{13}$ & [-1, 1] & [-1, 1] & [26, 1] & $D_{26}$ & $.2393$\\
$\frac{1}{13}$ & [1, -1] & [-1, 1] & [52, 4] & $D_{52}$ & $.2393$\\
$\frac{1}{13}$ & [-i, i] & [-i, i] & [52, 1] & $Z_{13} \rtimes Z_{4}$ & $.2393$\\
$\frac{1}{14}$ & [-1, 1] & [-1, 1] & [28, 3] & $D_{28}$ & $.2225$\\
$\frac{1}{14}$ & [-i, i] & [-1, 1] & [56, 4] & $Z_{4} \times D_{14}$ & $.2225$\\
$\frac{1}{14}$ & [1, -1] & [-1, 1] & [14, 1] & $D_{14}$ & $.2225$\\
$\frac{1}{14}$ & [-i, i] & [-i, i] & [28, 1] & $Z_{7} \rtimes Z_{4}$ & $.2225$\\
$\frac{1}{15}$ & [-1, 1] & [-1, 1] & [30, 3] & $D_{30}$ & $.2079$\\
$\frac{1}{15}$ & [1, -1] & [-1, 1] & [60, 12] & $D_{60}$ & $.2079$\\
$\frac{1}{15}$ & [-i, i] & [-i, i] & [60, 3] & $Z_{15} \rtimes Z_{4}$ & $.2079$\\
$\frac{2}{21}$ & [-1, 1] & [-1, 1] & [42, 5] & $D_{42}$ & $.2948$\\
$\frac{2}{23}$ & [-1, 1] & [-1, 1] & [46, 1] & $D_{46}$ & $.2698$\\
$\frac{2}{25}$ & [-1, 1] & [-1, 1] & [50, 1] & $D_{50}$ & $.2487$\\
$\frac{2}{27}$ & [-1, 1] & [-1, 1] & [54, 1] & $D_{54}$ & $.2306$\\
$\frac{2}{29}$ & [-1, 1] & [-1, 1] & [58, 1] & $D_{58}$ & $.2150$\\
$\frac{2}{31}$ & [-1, 1] & [-1, 1] & [62, 1] & $D_{62}$ & $.2013$\\
$\frac{3}{31}$ & [-1, 1] & [-1, 1] & [62, 1] & $D_{62}$ & $.2994$\\
$\frac{3}{32}$ & [-1, 1] & [-1, 1] & [64, 52] & $D_{64}$ & $.2903$\\
$\frac{3}{32}$ & [-i, i] & [-1, 1] & [64, 53] & $QD_{64}$ & $.2903$\\
$\frac{3}{32}$ & [-i, i] & [-i, i] & [64, 54] & $Q_{64}$ & $.2903$\\
$\frac{3}{34}$ & [-1, 1] & [-1, 1] & [68, 4] & $D_{68}$ & $.2737$\\
$\frac{3}{34}$ & [1, -1] & [-1, 1] & [34, 1] & $D_{34}$ & $.2737$\\
$\frac{3}{34}$ & [-i, i] & [-i, i] & [68, 1] & $Z_{17} \rtimes Z_{4}$ & $.2737$\\
$\frac{3}{35}$ & [-1, 1] & [-1, 1] & [70, 3] & $D_{70}$ & $.2660$\\
$\frac{3}{37}$ & [-1, 1] & [-1, 1] & [74, 1] & $D_{74}$ & $.2520$\\
$\frac{3}{38}$ & [1, -1] & [-1, 1] & [38, 1] & $D_{38}$ & $.2455$\\
$\frac{3}{46}$ & [1, -1] & [-1, 1] & [46, 1] & $D_{46}$ & $.2035$\\
\hline
\end{tabular}
\caption{
\label{TS_wide}
Flavour symmetries of $U_{CKM}^{LO}$, where $\mathcal{G}_{d} \sim Z_{m}$, $\mathcal{G}_{u} \sim Z_{n}$ with $m, n<5$ and $\mathcal{O}(\mathcal{G}_{\mathcal{Q}}) \leq 75$.  We display outcomes with distinct groups and $\sin \theta_{C}$ (for each case there were duplicates where different $T$ and $S$ generators from the ones shown result in the same group and same physical angle).}
\end{table}

\begin{table}
\begin{tabular}{|c|c|c|c|c|c|}
\hline 
 c & $T_{diag}$ & $S_{i}$ & GAP-ID & Group Structure & $\sin \theta_{C}$ \\
\hline 
$\frac{4}{55}$ & [-1, 1] & [-1, 1] & [110, 5] & $D_{110}$ & $.2265$\\
$\frac{4}{55}$ & [1, -1] & [-1, 1] & [220, 14] & $D_{220}$ & $.2265$\\
$\frac{5}{69}$ & [-1, 1] & [-1, 1] & [138, 3] & $D_{138}$ & $.2257$\\
$\frac{5}{69}$ & [1, -1] & [-1, 1] & [276, 9] & $D_{276}$ & $.2257$\\
$\frac{6}{83}$ & [-1, 1] & [-1, 1] & [166, 1] & $D_{166}$ & $.2252$\\
$\frac{6}{83}$ & [1, -1] & [-1, 1] & [332, 3] & $D_{332}$ & $.2252$\\
$\frac{7}{97}$ & [-1, 1] & [-1, 1] & [194, 1] & $D_{194}$ & $.2248$\\
$\frac{7}{97}$ & [1, -1] & [-1, 1] & [388, 4] & $D_{388}$ & $.2248$\\
\hline
\end{tabular}
\caption{
\label{TS_tight}
Flavour symmetries of $U^{LO}_{CKM}$, where $\mathcal{G}_{d} \sim Z_{m}$,  $\mathcal{G}_{u} \sim Z_{n}$ with $m,n \leq 3$ and $\mathcal{O}(\mathcal{G}_{\mathcal{Q}}) \leq 1000$. We display only outcomes with distinct groups and $\sin \theta_C$ (for each case there were duplicates where different $T$ and $S$ generators from the ones shown result in the same group and same physical angle).}
\end{table}

\subsection{\boldmath{$G_{d} \sim Z_{m1}^{d} \times Z_{m2}^{d}$, \, $G_{u} \sim Z_{n}^{u}$}}
\label{subsec:udd}

We now enlarge the symmetry assignment in the down sector by allowing a direct product of cyclic groups, in analogy to the $Z_{2} \times Z_{2}$ symmetry of the Majorana neutrino mass matrix. We again allow $a,b \in \lbrace 0,1 ...50 \rbrace$, $\mathcal{O}(T,S_{1},S_{2}) \leq 4$, and $\mathcal{O}(\mathcal{G}_{\mathcal{Q}}) \leq 75$, but restrict the Cabibbo window to $.2 \leq \sin \theta_{C} \leq .24$ to obtain predictions closer to the experimental value. The results are presented in Table \ref{TSS}, where we see that the only new group found in comparison to Table \ref{TS_wide} is $Z_{3} \times D_{14}$, which also predicts $\sin \theta_{C} \simeq .2225$.  The result is highlighted in blue because the eigenvalues of $S_{i1}$ are clearly degenerate, and therefore not capable of distinguishing the two generations.\footnote{In creating Table \ref{TSS} we have otherwise filtered results where generators carry degenerate eigenvalues.  Every group present in the table could have also been generated by these physically uninteresting matrices.}   In any event, from the model-building perspective, this group is also not an interesting result as it does no more work than $D_{14}$.   

It is of course not surprising that we do not find any new quantizations of $\sin \theta_{C}$, as this is totally controlled by the range in $a$/$b$ scanned and the Cabibbo window, which were chosen to be the same as (or contained within) those used for Table \ref{TS_wide}.  It is also not concerning that, for example, $D_{28}$ is `generated' by three matrices when it is well known that Dihedrals can be closed with only two.  After all, a finite group $\mathcal{G}_{\mathcal{F}}$ can be `generated' by as many as $\mathcal{O}(\mathcal{G}_{\mathcal{F}})$ elements!  So, when we say that Dihedrals have two generators, we mean that the \emph{smallest} set of generating elements for Dihedral groups is $\mathcal{O}(2)$.  Indeed, due to the internal ordering of group elements, if one asks GAP for the generators $f_{i}$ of $SmallGroup(28,3)$ corresponding to $D_{28}$, a three element set is returned\footnote{Even Abelian groups like $Z_{4}$ will sometimes return multi-element generator sets with the $GeneratorsOfGroup$ command.}:
\begin{equation}
GeneratorsOfGroup(SmallGroup(28,3)) \,\,\,=\,\,\,\left[ f_{1}, f_{2}, f_{3} \right]
\end{equation}
However, GAP also knows that there is a smaller subset of these three `generators' that will also do the job:
\begin{equation}
MinimalGeneratingSet(SmallGroup(28,3)) \,\,\,=\,\,\,\left[ f_{1}, f_{2} \cdot f_{3}\right]
\end{equation}
The very same reasoning can also be applied in reverse to Table \ref{TS_wide}, where the group $\Psi(6,2)$ would normally be assigned three generators to better reveal its structure in terms of three cyclic symmetries ($(Z_{6} \times Z_{2}) \rtimes Z_{2}$), yet can in fact be generated by two.  $\Delta(27) \in \Delta(3N^{2})$ $((Z_{3} \times Z_{3}) \rtimes Z_{3}$), a popular group for model-building in the leptonic sector \cite{deMedeirosVarzielas:2006fc, Ma:2006ip, Varzielas:2015aua}, is a well known example of this.

\begin{table}
\begin{tabular}{|c|c|c|c|c|c|c|}
\hline 
 c & $T_{diag}$ & $S_{i1}$ & $S_{i2}$ & GAP-ID & Group Structure & $\sin \theta_{C}$ \\
\hline 
$\frac{1}{13}$ & [-1, 1] & [-1, 1] & [1, -1] & [52, 4] & $D_{52}$ & $.2393$\\
$\frac{1}{13}$ & [-i, i] & [-i, i] & [i, -i] & [52, 1] & $Z_{13} \rtimes Z_{4}$ & $.2393$\\
$\frac{1}{14}$ & [-1, 1] & [-1, 1] & [1, -1] & [28, 3] & $D_{28}$ & $.2225$\\
$\frac{1}{14}$ & [-1, 1] & [-1, 1] & [-i, i] & [56, 4] & $Z_{4} \times D_{14}$ & $.2225$\\
$\frac{1}{14}$ & [-i, i] & [-i, i] & [i, -i] & [28, 1] & $Z_{7} \rtimes Z_{4}$ & $.2225$\\
\ras{$\frac{1}{14}$} & \ras{[1, -1]} & \ras{[E(3)$^2$, E(3)$^2$]} & \ras{[-1, 1]} & \ras{[42, 4]} & \ras{$Z_{3} \times D_{14}$} & \ras{$.2225$}\\
$\frac{1}{15}$ & [-1, 1] & [-1, 1] & [1, -1] & [60, 12] & $D_{60}$ & $.2079$\\
$\frac{1}{15}$ & [-i, i] & [-i, i] & [i, -i] & [60, 3] & $Z_{15} \rtimes Z_{4}$ & $.2079$\\
\hline
\end{tabular}
\caption{
\label{TSS}
Flavour symmetries of $U_{CKM}^{LO}$, where $\mathcal{G}_{d} \sim Z_{m_{1}} \times Z_{m_{2}}$, $\mathcal{G}_{u} \sim Z_{n}$ with $m, n<5$, $\mathcal{O}(T, S) < 5$ and $\mathcal{O}(\mathcal{G}_{\mathcal{Q}}) \leq 75$. We display outcomes with distinct groups and $\sin \theta_{C}$ (for each case there were duplicates where different $T$ and $S$ generators from the ones shown result in the same group and same physical angle).}
\end{table}

\subsection{\boldmath{$G_{d} \sim Z_{m}^{d}$, \, $G_{u} \sim Z_{n1}^{u} \times Z_{n2}^{u}$}}
\label{subsec:uud}

We also naively scan the symmetry assignment corresponding to two up-sector residual generators, as opposed to two (non-diagonal) down-sector generators.  Utilizing the same parameter ranges as in Section \ref{subsec:udd}, we find the exact same results as those presented in Table \ref{TSS}, with $T \leftrightarrow S$.  This result is unsurprising, as any physical symmetry must be basis independent, and moving between the two symmetry assignments in Sections \ref{subsec:udd} and \ref{subsec:uud} requires nothing more than a basis transformation.  To see this, simultaneously rotate the 3 generators of Section \ref{subsec:udd} with the inverse of the operation in Eq. (\ref{eq:SD}) (where we implicitly chose a basis to work in):
\begin{equation}
\lbrace S_{D}, T_{1}, T_{2} \rbrace \,\,\, \longrightarrow \,\,\, U_{CKM}^{\dagger} \lbrace S_{D}, T_{1}, T_{2} \rbrace U_{CKM}\,\equiv\, \lbrace S, T_{D1}, T_{D2} \rbrace
\end{equation}
where $T_{D1}$ and $T_{D2}$ are non-diagonal generators analogous to $S_{Di}$ given in Eq. (\ref{eq:SD}).  However, we are of course entirely free to relabel our generators; $T_{l}$ and $S_{i}$ are both diagonal matrices sourced from equivalent lists of all possible charge permutations in Eq. (\ref{eq:T}): 
\begin{equation}
\lbrace S, T_{D1}, T_{D2} \rbrace \,\,\, \underset{T \leftrightarrow S}{\longrightarrow} \,\,\, \lbrace T, S_{D1}, S_{D2} \rbrace
\end{equation}
We have now arrived at the generator set for the symmetry assignment in Section \ref{subsec:udd}. 

\subsection{\boldmath{$G_{d} \sim Z_{m1}^{d} \times Z_{m2}^{d} $, \, $G_{u} \sim Z_{n1}^{u} \times Z_{n2}^{u}$}}

As a final check, we also scan the symmetry assignment where two generators are assigned to the up and down sector.  We keep the same input parameters as in Section \ref{subsec:udd}.  Although more groups are closed (given the larger number of generators), after excluding the redundant cases (with the same angle and same $\mathcal{G}_{\mathcal{Q}}$) the results are again the same as in the previous two sections --- the additional generator in either the up or down sector does no work for us, at least within the parameter ranges we choose.

\subsection{Looking for broken symmetries --- a consistency check}
The groups we find are sourced from the explicit representation of the residual generators, Eq. (\ref{eq:T}) and Eq. (\ref{eq:SD}).  The method is ignorant of what these matrices actually represent, i.e. the symmetry assignments of the physical Lagrangian.  Hence, from a completely agnostic perspective, we might also use the bottom-up method to analyze the generator associated with the upper $2 \times 2$ sub-matrix of the Wolfenstein parameterization by expanding Eq. (\ref{eq:LOCKM}) about the Cabibbo angle:
\\
\begin{equation}
\label{eq:SD2}
S^{\lambda}_{Di}=
\renewcommand{\arraystretch}{1.5}
\left(
\begin{array}{cc}
e^{i\alpha_{2i}} \lambda^{2} + e^{i\alpha_{1i}} (\frac{\lambda^{2}}{2} -1)^{2} & \left(e^{i\alpha_{1i}}-e^{i\alpha_{2i}}\right) (\frac{\lambda^{3}}{2}-\lambda)\\
 \left(e^{i\alpha_{1i}}-e^{i\alpha_{2i}}\right) (\frac{\lambda^{3}}{2}-\lambda) &e^{i\alpha_{1i}} \lambda^{2} + e^{i\alpha_{2i}} (\frac{\lambda^{2}}{2} -1)^{2} 
\end{array}
\right)
\end{equation}
\\
While this generator reflects a trivial rewriting of the original mixing matrix and only changes the numerical values of its elements by small amounts (for substantially small $\lambda$), it is a priori entirely plausible that the (exact) structures of Eq. (\ref{eq:SD}) and Eq. (\ref{eq:SD2}) for a given quantized value of $\theta_{C}/\lambda$ generate different parent groups $\mathcal{G}_{\mathcal{Q}}$ when closed with $T$.  That is, minor numerical shifts of $|V^{LO}_{ij}|$ might be sourced by entirely different group structures. 

 \emph{However}, Eq. (\ref{eq:SD2}) reflects quark mixing that is only unitary up to $\mathcal{O}(\lambda^{4})$:
\begin{equation}
V_{\lambda} V^{\dag}_{\lambda}=
\renewcommand{\arraystretch}{1.25}
\left(
\begin{array}{cc}
1 + \mathcal{O}(\lambda^{4}) & 0\\
0 & 1 + \mathcal{O}(\lambda^{4})
\end{array}
\right)
\end{equation}
\\
and hence does not generate a symmetry of the Lagrangian.  One might then be tempted to interpret it as a `broken-symmetry' generator.  Regardless, we would not expect such a matrix to actually close a finite mathematical group, as the generator itself should not be of finite order, $\mathcal{O}(S^{\lambda}_{Di}) = \infty$.  Indeed, upon running our scripts with Eq. (\ref{eq:T}) and Eq. ({\ref{eq:SD2}) as the potential group generators, we find that no NA finite flavour groups are closed.

\section{General trends and limitations of the bottom-up technique}
\label{sec:Limits}
While the bottom-up technique described above is a powerful tool that can be used to rapidly identify viable NA discrete symmetries useful for model-building, we here discuss some of its limitations.  Regarding physics, the method only applies to direct and semi-direct models, respectively those that either predict all angles in the mixing matrix (in this case leaves no freedom in the 2x2 submatrix) or to those models that predict a column of the mixing matrix (like tri-maximal mixing matrices in the case of leptons \cite{Talbert:2014bda}).  The method does \emph{not} apply to cases where the specific residual symmetries are not subgroups of the actual flavour symmetry of the model (referred to as indirect models \cite{King:2013eh}).

The method is also sensitive to the user-defined input parameters, including the scan ranges for the various $a$/$b$ (related to the discretization of the phases and $\theta_{C}$), the allowed quantization range for $\sin\theta_{C}$, the maximum allowed order for $\mathcal{G}_{u/d}$, and the maximum allowed order for $\mathcal{G}_{\mathcal{F}}$.  Widening or increasing any of these parameters quickly produces many more group closures, and hence also slows operations.  Figure \ref{Plots} plots an independent variation of each of these four `tunes' (given the symmetry assignment in Section \ref{subsec:ud}) against the number of finite, NA groups closed (all closed groups are displayed, whether they are duplications or not).  These plots are meant as a qualitative illustration of the growth of group closures.  We see that increasing the scan ranges of $a$/$b$ ($a_{\text{max}}$ - allowing for a finer gridding of $\theta_{C}$) and widening the allowed range of $\sin \theta_{C}$ produces a roughly linear increase in group closures, whereas increasing the allowed order of the parent symmetry eventually plateaus (Figure \ref{Plots}D).  This plateau is sensible;  there will only be a limited number of finite groups closed when all constraints are also finite.  Had we increased the value of $a_{\text{max}}$ to 35 in Figure \ref{Plots}D, for example, the plateau would occur at 120 groups for $MaxOrder (\mathcal{G}_{\mathcal{F}}) \geq 170$.  

Figure \ref{Plots}C, on the other hand, also exhibits an overall plateau in group closures despite an unrestricted $\mathcal{O}(\mathcal{G}_{\mathcal{F}})$.  This behavior is less intuitive, though clearly an artifact of our constraint on $\sin\theta_{C}$. This can be seen from Table \ref{largeangle} where generators of order 3 appear even though the prediction of the Cabibbo angle is un-physical. Therefore, relaxing the Cabibbo window would increase the number of groups closed from $\mathcal{O}(T,S) \leq 2$ and $\mathcal{O}(T,S) \leq 3$ through the method. To confirm that the plateau exists when $.2 \leq \sin\theta_{C} \leq .3$, we also ran two other scans where the effective number of values for the Cabibbo angle, $\theta_C$, encoded by rational parameter $c$, are reduced to four and one (there are 10 active Cabibbo angles in Figure \ref{Plots}c).  In both instances we see plateaus beginning at $\mathcal{O}(T,S) \leq 2$ and $\mathcal{O}(T,S) \leq 4$, and in the single-$c$ scan the final plateau remains up to  $\mathcal{O}(T,S) \leq 8$ (we only ran up to $\mathcal{O}(T,S) \leq 7$ for the four-$c$ scan).  Intriguingly, there are plateaus at $4^{1} \times (\# \,\,of\,\, c's)$ and $4^{2} \times (\# \,\,of\,\, c's)$ in all three scans.  So, there are plateaus at 4 and 16 group closures for one active $c$, 16 and 64 group closures for four active $c$'s, and 40 and 160 group closures for 10 active $c$'s.  We have checked that there are (as must be the case) more closures of \emph{Abelian} finite groups as $\mathcal{O}(T,S)$ increases, but not the NA groups that we are interested in.  

Another difficulty (not necessarily associated to the bottom-up technique) arises when the angles considered are very small, as the order of the predictive group then increases significantly as do the computational weights of the associated GAP objects.  This correlation between group order and associated mixing angle is illustrated clearly in Table \ref{TS_tight}, where the increase in precision of the predicted angles came with an associated increase in the order of the groups. This is easy to understand for Dihedral groups due to the associated geometric interpretation in terms of polygons. It follows then that to obtain a small angle, one naturally needs to have a respectively small $c$ parameter. For example, to quantize the smallest quark mixing angle - $\theta^{q}_{13} \approx \pi/900$ - we obtain Dihedral groups with $Order(\mathcal{G_F}) \gtrsim \mathcal{O}(1000)$. This can then be treated as a lower bound on the order of groups necessary to quantize the full CKM matrix, and all associated degrees of freedom. Indeed, we performed a short, dedicated ``bottom-up'' scan to hunt for discrete symmetries capable of quantizing the full CKM matrix, yet do not find any groups identifiable by GAP and its \emph{Small Groups} Library, given our parameter ranges and computational expense. These results are consistent with previous studies and with the naive estimate of the necessary order of the predictive group given above -- if a relevant finite group had an order smaller than $\mathcal{O}(1000)$, our approach should also be able to find it.

\begin{figure}[t]
\centering
\includegraphics[scale=.4]{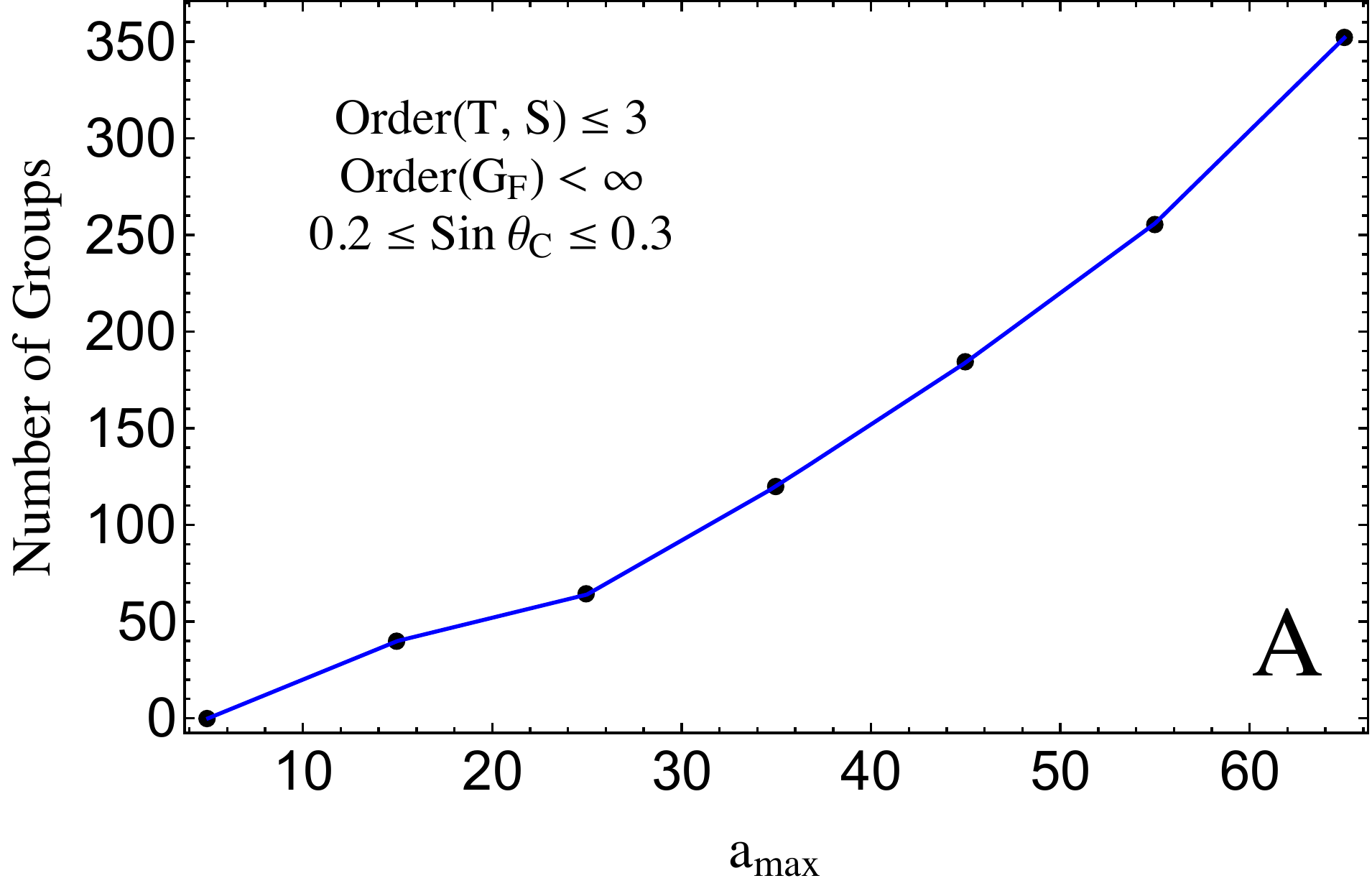}
\vspace*{0.75cm}
\hspace*{1.0cm}
\includegraphics[scale=.4]{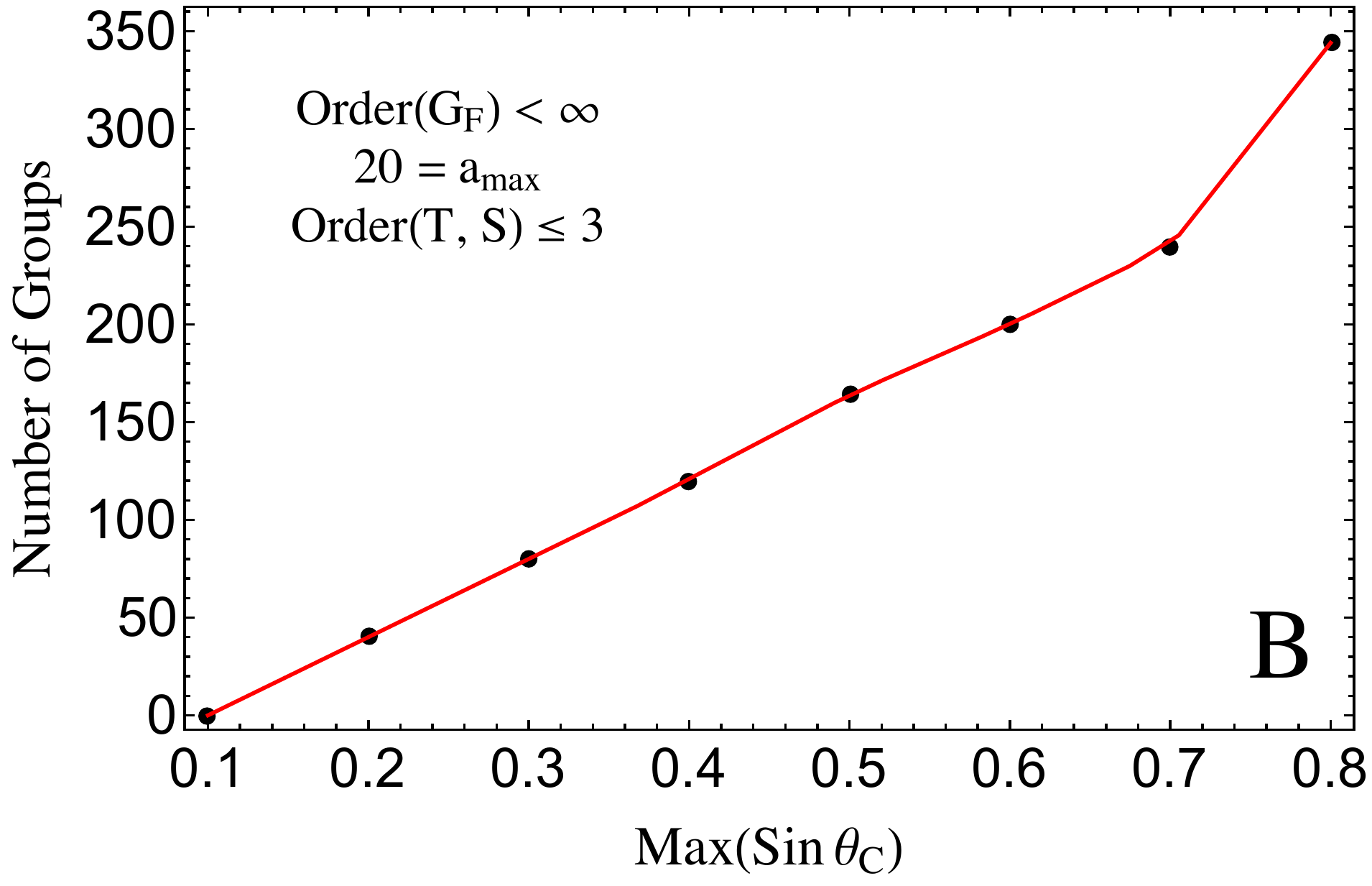}
\includegraphics[scale=.4]{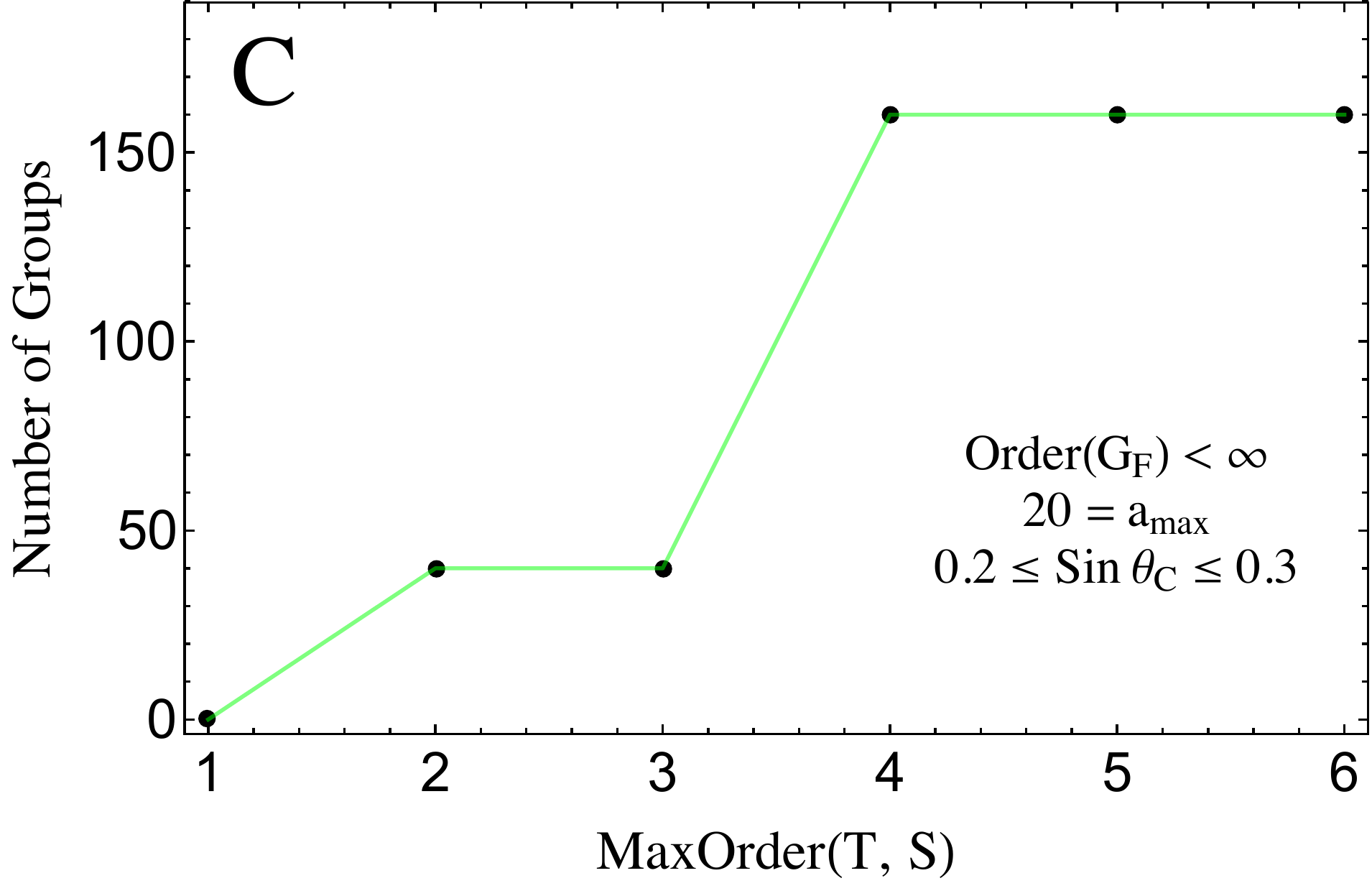}
\hspace*{1.0cm}
\includegraphics[scale=.4]{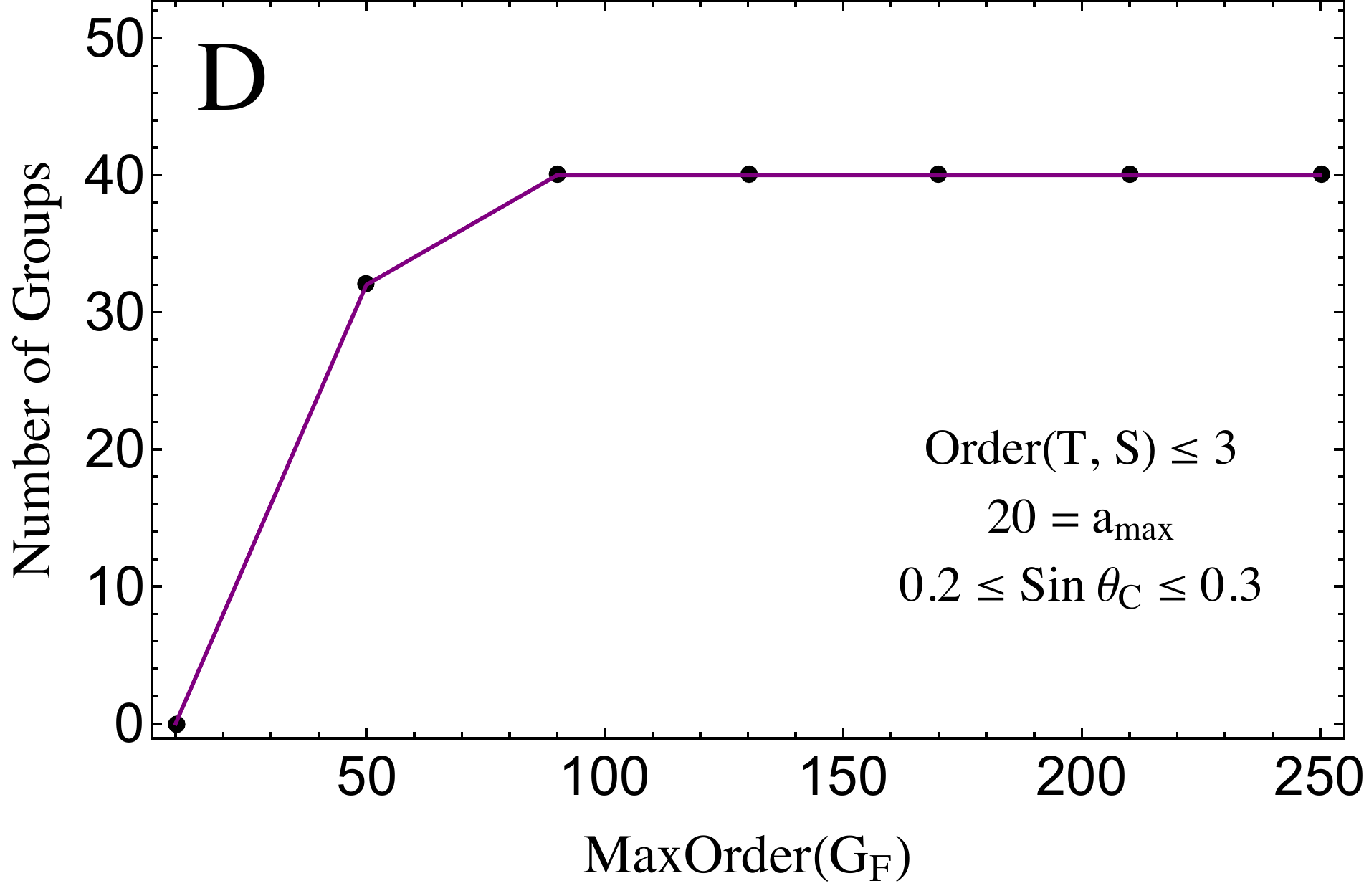}
\caption{\label{Plots} Tables showing the number of parent groups $\mathcal{G}_{\mathcal{F}}$ found when varying four inputs to the `bottom-up' approach, namely the discretization parameters $a$ and $b$ (A), the allowed quantization range of $\sin \theta_{C}$ (B), the maximum allowed order of the residual Abelian symmetry groups $\mathcal{O}(\mathcal{G}_{u}, \mathcal{G}_{d})$ (C), and the maximum allowed order of the parent NA symmetry group $\mathcal{O}(\mathcal{G}_{\mathcal{F}})$ (D).  In each case the other 3 inputs are left fixed to the values shown in the tables.  The number of groups given represents the number of raw groups closed by the method, and does not include any trimming of charge degeneracies, etc.  The curves represent first-order interpolations of the data, and are present as a visual aid only --- they do not represent any theory.}
\end{figure}

\section{Conclusions \label{sec:conclusions}}
We have applied the bottom-up [re]construction procedure of \cite{Talbert:2014bda} to scan over possible NA finite groups $\mathcal{G}_{\mathcal{Q}}$ capable of quantizing the Cabibbo angle of CKM mixing.  This study complements other `top-down' scans which, by virtue of the restrictions put on the irreducible representations of the parent symmetry or other theory biases (e.g. searching for groups that also work for the leptons), do not find or otherwise obscure interesting small groups that can do the same job.  After all, no group has been found that can fully quantize Eq. (\ref{eq:num}), and theorists interested in using NA finite groups in the quark sector should therefore consider the possibility that such symmetries, if natural, may make predictions that are substantially corrected via other mechanisms.  

Our scans find multiple candidate groups for $\mathcal{G}_{\mathcal{Q}}$ in Tables \ref{TS_wide}-\ref{TSS} , including small semi-direct product structures like $Z_{7} \rtimes Z_{4}$ and $Z_{3} \rtimes Q_{8}$, $\Psi(6,2)$, and (Quasi)Dihedrals.  Given more liberal adjustments of the input parameters, our scans also find other groups like $S_{3}$ and $\Sigma(32)$ found in Table \ref{largeangle}.  Our results seem consistent with former studies of quark mixing, modulo our starting point of 2 dimensional representations for residual generators in the up and down sector.  For larger groups of $\mathcal{O}(10^{2})$ we can reproduce the PDG values for the (12) and (21) matrix elements of $U^{PDG}_{CKM}$.  We thus also validate the [re]construction procedure, which may be of further use model-building both within Standard Model and BSM mixing scenarios.  

\section{Acknowledgments}

This project is supported by the European Union's Seventh Framework Programme for research, technological development and demonstration under grant agreement no PIEF-GA-2012-327195 SIFT.  RWR appreciates the discussion and helpful comments from Walter Winter.  JT is grateful to the University of Southampton, where collaboration on this project began, and to Prof. G.G. Ross for ongoing discussions and encouragement.  JT acknowledges support from the Senior Scholarship Trust of Hertford College, University of Oxford.

\appendix

\section{\label{App} Symmetries for other angles}

In this appendix we include results for a short scan where we constrain the Cabibbo window to $.7 \leq \sin \theta_{C} \leq .8$.  Our purpose is to illustrate that our scripts, given appropriate inputs, can in fact find groups that may be naively expected given the generator representations in Eqs. (\ref{eq:T}) and (\ref{eq:SD}).  Results for the symmetry assignment $\mathcal{G}_{u/d} \sim Z_{n/m}$ are found in Table \ref{largeangle}, where we have input $a,b \in \lbrace 0,1 ...10 \rbrace$, $\mathcal{O}(T,S) \leq 4$, and $\mathcal{O}(\mathcal{G}_{\mathcal{Q}}) \leq 75$.  We see that groups like $S_3$\footnote{Note the diagonal generator is the order 3 generator with powers of E(3), i.e. it is not in the basis where the generators represent the geometrical symmetries of the triangle.} and $\Sigma(2N^2)$ ($(Z_4 \times Z_4) \rtimes Z_2$) are generated as expected, which are known in the literature to generate lepton mixing angles (see e.g. \cite{Hernandez:2015dga} and references therein). Additionally, Table \ref{largeangle} also provides evidence for the plateau seen in Figure \ref{Plots}C is an artifact of our constraint on $\sin\theta_{C}$. If this were relaxed, the plateau would disappear.

\begin{table}[tp]
\makebox[\textwidth]{
\begin{tabular}{|c|c|c|c|c|c|}
\hline 
 c & $T_{diag}$ & $S_{i}$ & GAP-ID & Group Structure & $\sin \theta_{C}$ \\
\hline 
$\frac{1}{4}$ & [-1, 1] & [-1, 1] & [8, 3] & $D_{8}$ & $.7071$\\
$\frac{1}{4}$ & [E(3)$^2$, 1] & [-1, 1] & [18, 3] & $Z_{3} \times S_{3}$ & $.7071$\\
$\frac{1}{4}$ & [-i, 1] & [-1, 1] & [32, 11] & $\Sigma(2 \cdot 4^{2})$ & $.7071$\\
$\frac{1}{4}$ & [E(3)$^2$, E(3)] & [-1, 1] & [6, 1] & $S_{3}$ & $.7071$\\
$\frac{1}{4}$ & [-i, i] & [E(3)$^2$, 1] & [36, 6] & $Z_{3} \times (Z_{3} \rtimes Z_{4})$ & $.7071$\\
$\frac{1}{4}$ & [-i, i] & [E(3)$^2$, E(3)] & [12, 1] & $Z_{3} \rtimes Z_{4}$ & $.7071$\\
$\frac{1}{4}$ & [-i, i] & [-i, i] & [8, 4] & $Q_{8}$ & $.7071$\\
$\frac{2}{7}$ & [-1, 1] & [-1, 1] & [14, 1] & $D_{14}$ & $.7818$\\
$\frac{2}{7}$ & [-i, i] & [-1, 1] & [56, 4] & $Z_{4} \times D_{14}$ & $.7818$\\
$\frac{2}{7}$ & [1, -1] & [-1, 1] & [28, 3] & $D_{28}$ & $.7818$\\
$\frac{2}{7}$ & [-i, i] & [-i, i] & [28, 1] & $Z_{7} \rtimes Z_{4}$ & $.7818$\\
\hline
\end{tabular}}
\caption{\label{largeangle} Flavour symmetries of $U_{CKM}^{LO}$ where $\mathcal{G}_{u/d} \sim Z_{n/m}$ with $\mathcal{O}(T,S) \leq 4$ and $\mathcal{O}(\mathcal{G}_{F}) \leq 75$. We have searched the (non-physical) range $.7 \leq \sin \theta_{C} \leq .8$.}
\end{table}

\end{document}